\documentclass[epjST]{svjour}
 
\usepackage{graphics}
\usepackage{eurosym}
\usepackage{amsmath}
\usepackage{hyperref}
 
\begin{document}
 
\title{Reconfigurable computing for Monte Carlo simulations:
results and prospects of the Janus project}

\author{
M.~Baity-Jesi\inst{1,2}\and
R.~A.~Ba\~nos\inst{3,2} \and
A.~Cruz\inst{3,2} \and
L.~A.~Fernandez\inst{1,2} \and
J.~M.~Gil-Narvion\inst{2} \and
A.~Gordillo-Guerrero\inst{4,2}
M.~Guidetti\inst{2} \and
D.~I\~niguez\inst{5,2} \and
A.~Maiorano\inst{6,2} \and
F.~Mantovani\inst{7}\footnote{Present address: Physics Department,
University of Regensburg, Germany.} \and
E.~Marinari\inst{8} \and
V.~Martin-Mayor\inst{1,2} \and
J.~Monforte-Garcia\inst{3,2} \and
A.~Mu\~noz~Sudupe\inst{1} \and
D.~Navarro\inst{9} \and
G.~Parisi\inst{8} \and
M.~Pivanti\inst{6} \and
S.~Perez-Gaviro\inst{2} \and
F.~Ricci-Tersenghi\inst{8} \and
J.~J.~Ruiz-Lorenzo\inst{10,2} \and
S.~F.~Schifano\inst{11} \and
B.~Seoane\inst{1,2} \and
A.~Tarancon\inst{3,2} \and
P.~Tellez\inst{3}\and
R.~Tripiccione\inst{7} \and
D.~Yllanes\inst{6,2}
}

\institute{Departamento de F\'{\i}sica Te\'orica I, Universidad Complutense, 28040 Madrid, Spain. \and
Instituto de Biocomputaci\'on y F{\'{\i}}sica de Sistemas Complejos (BIFI), Zaragoza, Spain. \and
Departamento de F\'{\i}sica Te\'orica, Universidad de Zaragoza, 50009 Zaragoza, Spain. \and 
Departamento de Ingenier\'{\i}a El\'ectrica, Electr\'onica y Autom\'atica, Universidad de Extremadura,  10071. C\'aceres, Spain.\and
Fundaci\'on ARAID, Diputaci\'on General de Arag\'on, Zaragoza, Spain.\and
Dipartimento di Fisica, La Sapienza Universit\`a di Roma, 00185 Rome, Italy. \and
Dipartimento di Fisica Universit\`a di Ferrara and INFN - Sezione di Ferrara, Ferrara, Italy. \and
Dipartimento di Fisica, IPCF-CNR, UOS Roma Kerberos and INFN,
La Sapienza Universit\`a di Roma, 00185 Rome, Italy\and
Departamento de Ingenier\'{\i}a, Electr\'onica y Comunicaciones and Instituto de Investigaci\'on en Ingenier\'{\i}a de Arag\'on (I3A), Universidad de Zaragoza, 50018 Zaragoza, Spain. \and
Departamento de F\'{\i}sica, Universidad de Extremadura, 06071 Badajoz, Spain. \and
Dipartimento di Matematica e Informatica Universit\`a di Ferrara and INFN - Sezione di Ferrara, Ferrara, Italy.
}

\abstract{
We describe Janus, a massively parallel FPGA-based computer optimized for the
simulation of spin glasses, theoretical models for the behavior of glassy
materials. FPGAs (as compared to GPUs or many-core processors) provide a
complementary approach to massively parallel computing. In particular, our
model problem is formulated in terms of binary variables, and floating-point
operations can be (almost) completely avoided. The FPGA architecture allows us
to run many independent threads with almost no latencies in memory access, thus
updating  up to 1024 spins per cycle. 
We describe Janus in detail and we summarize the physics results obtained in
four years of operation of this machine; we discuss two types of physics
applications: long  simulations on very large systems (which try to mimic and
provide understanding about the experimental non-equilibrium dynamics), and
low-temperature equilibrium simulations using an artificial parallel tempering
dynamics. The time scale of our non-equilibrium simulations spans eleven orders
of magnitude (from picoseconds to a tenth of a second). On the other hand, our
equilibrium simulations are unprecedented both because of the low temperatures
reached and for the large systems that we have brought to equilibrium. A
finite-time scaling ansatz emerges from the  detailed comparison of the two sets
of simulations.
Janus has made it possible to perform spin-glass simulations that would take
several decades on more conventional architectures. The paper ends with an
assessment of the potential of possible future versions of the Janus
architecture, based on state-of-the-art technology.
} 
 
\maketitle
 
\section{Overview}
\label{sec:intro}
A major challenge in condensed-matter physics is the understanding of
glassy behavior~\cite{INTROGLASS}. Glasses are materials of the greatest
industrial relevance (aviation, pharmaceuticals, automotive, etc.) that do
not reach thermal equilibrium in human lifetimes. Important material
properties, such as the compliance modulus or the specific heat,
significantly depend on time even if the material is kept for months (or
years) under constant experimental conditions~\cite{STRUICK}. This
sluggish dynamics is a major problem for the experimental and theoretical
investigation of glassy behavior, placing numerical simulations at the
center of the stage.

Spin glasses are the prototypical glassy systems most widely studied
theoretically~\cite{MYDOSH,YOUNG}. Simulating spin glasses poses a
formidable challenge to state-of-the-art computers: studying the
surprisingly complex behavior governed by deceivingly simple dynamical
equations still requires inordinately large computing resources. In a
typical spin-glass model, the dynamical variables, named spins, are
discrete and sit at the nodes of discrete $D$-dimensional lattices.  In
order to make contact with experiments, we need to follow the evolution of
a large enough 3D lattice, say $80^3$ sites, for time periods of the order
of $1$ second. One Monte Carlo step ---the update of all the $80^3$ spins
in the lattice---  roughly corresponds to $10^{-12}$ seconds, so we need
some $10^{12}$ such steps, that is $\sim10^{18}$ spin updates.
Furthermore, in order to account for the disorder we have to collect
statistics on several ($\sim 10^2$) copies of the system, adding up to
$\sim 10^{20}$ Monte Carlo spin updates. It is essential to realize that
the correct study of glassy behavior requires that the Monte Carlo steps
for each copy be {\em consecutive}.  Therefore, performing this simulation
program in a reasonable time frame (say, less than one year) requires a
computer system able to update on average one spin in $1$ picosecond or
less. In conclusion, realistic simulations of spin glasses have been a
major computational challenge, which has been solved for the first time
only three or four years ago; the Janus project ---started in late 2006
and reviewed in this paper--- has played a major role in reaching this
goal.

A large amount of potential parallelism is easily identified in the Monte
Carlo algorithms appropriate for spin-glass simulations, but exploiting a
large fraction thereof is not easy. Indeed, at the time the Janus project
started, state-of-the-art simulation programs running on state-of-the-art
computer architectures were only able to exploit a tiny fraction of the
available parallelism, and had an average update time of one spin every
$\sim 1$ ns, meaning that the simulation campaign outlined above would
proceed for centuries. Since 2006, computer architectures have
consistently evolved towards wider and wider parallelization: many-core
processors with $\mathcal O(10)$ cores are now widely available and Graphics
Processing Units (GPUs) now have hundreds of what can be regarded as
``slim'' cores.  Careful optimization for these architectures indeed
increases performance by about one order of magnitude (slightly better
than one would predict according to Moore's law, see later for a detailed
analysis), but standard commercial computers are still not a satisfactory
option for large-scale spin-glass studies.

What is really needed is an architecture able to exploit all (or at least
a large fraction of) the parallelism associated to the Monte Carlo
evolution of one copy of the system. This basically means handling at the
same time spin variables sitting at non-neighbor sites of the lattice.
This is in principle possible within reasonable limits of hardware
complexity, since the associated operations are rather simple and control
can be shared across collaborating data paths; however, once this approach
is put in practice, a dramatic memory access bottleneck follows, which
must be circumvented by appropriate memory organization and access
strategies.  If this approach is successful (allowing for a degree of
parallelism of order $1000 \times$) then a further factor ($100 \times$)
can be compounded by trivially processing in parallel a corresponding
number of copies of the system.

The architectural requirements outlined above are at variance with those
targeted by off-the-shelf CPUs, so one can expect huge performance gains
from an application-oriented architecture. Application-oriented systems
have been used in several cases in computational physics, in the area of
spin-system simulations \cite{ogielski} but also in Lattice QCD
\cite{qcdoc,ape,qpace}  and for the simulation of gravitationally coupled
systems \cite{grape}. 

In this paper we describe Janus\footnote{From the name of the ancient
Roman god of doors and \emph{gates}.}, yet another application-driven
massively parallel processing system able to handle Monte Carlo
simulations of spin glasses.  Janus is based on  FPGA technology.  FPGAs
(field-programmable gate arrays) are integrated circuits that can be
configured at will after they have been assembled in an electronic system.
FPGAs are slow w.r.t standard processors and trade flexibility for
complexity (under general terms, any system of $n$ gates can be emulated
by an FPGA of complexity $n \log_2 n$, but $\log_2 n$ is a ``large''
number when $n \simeq 10^6 - 10^7$).  These disadvantages are offset by
the more dramatic speedup factors allowed by architectural flexibility.
Also, the development time of an FPGA-based system is short. Early
attempts in this direction were made several years ago within the SUE
project \cite{sueMachine}.  A conceptually similar approach would be to
consider an ASIC (application-specific integrated circuit), a custom-built
integrated circuit, which would further boost performance, at the price of
much larger development time and cost, and much less flexibility in the
design.

Our paper is organized as follows: in section 2 we describe the physics systems
that we want to simulate, elaborating on their relevance both in physics and
engineering; section 3 provides details on the Monte Carlo simulation approach
used in our work; section 4 describes the Janus architecture and its
implementation.  Section 5 summarizes the main physics results obtained after
four years of continuous operation of the machine.  Section 6
assesses the performance of Janus on our spin-glass simulations, using several
metrics, and compares with more standard solutions. We consider both those
technologies that were  available when Janus was developed and commissioned and
those that have been developed since the beginning of the project ($\approx 5$
years ago).  We also briefly discuss the performance improvements that may be
expected if one re-engineers Janus on the basis of the
technology available today. Our conclusions and outlook are in section 7.

\section{Spin glasses}
\label{sec:spin-glasses}
\begin{figure}
\begin{center}
\resizebox{0.75\columnwidth}{!}{\includegraphics{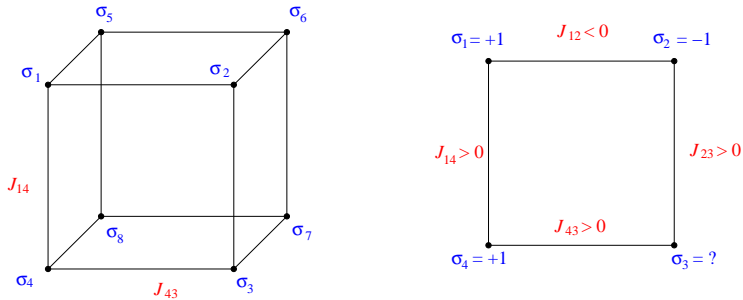} }
\caption{{\bf Left:} The spin-glass crystal lattice is obtained by
  periodically repeating the  unit cell in all directions.
  Dynamical variables, the spins (blue), take values $\sigma_i=\pm
  1$. Pairs of neighboring spins have either a tendency to
  take the same value or the opposite one. This is governed by a
  coupling constant (red) placed on the lattice link joining the two
  spins. As an example, $J_{43}>0$ would favor values of $\sigma_3, \sigma_4$
  such that $\sigma_3\cdot\sigma_4>0$.
  {\bf Right:} (frustration example) 
  Consider the front plaquette of the cell at left, and the
  given values of the coupling constants. Try to find an
  assignment of the four spins such that all four links are
  satisfied; set $\sigma_1=+1$ and find the correct assignment
  for $\sigma_3$ going around the plaquette clockwise: since
  $J_{12}<0$ we assign $\sigma_2=-1$ and then ($J_{23}>0$)
  $\sigma_3=-1$. On the other hand, going anticlockwise, we find that
  $\sigma_4=+1$, because $J_{14}>0$ and, since $J_{43}>0\,$, also
  $\sigma_3=+1\/$! If we change our initial guess to $\sigma_1=-1$,
  we also reach a contradictory assignments for $\sigma_3$.}
\label{FIG-BASICA}	    
\end{center}
\end{figure}

The basic ingredients of a spin glass (SG) are frustration and randomness
(see Fig.~\ref{FIG-BASICA}). One typical example is a metal in which we
replace some of its metallic atoms with magnetic ones.  We can roughly
describe its physics: the dynamical variables, the spins, represent atomic
magnetic moments, which interact via electrons in the conduction band
of the metal, inducing in this way an effective interaction which changes
in sign (the RKKY interaction) depending on the spatial location.  In some
materials, it is easy for the magnetic moments to lie in only one direction
(and not in the original three-dimensional space) and we can consider that
they take only two values. This is what we call an Ising material (while
a three-dimensional version would be a Heisenberg spin glass).
Finally we can assume that these spins sit at
the nodes of a crystal lattice (Fig.~\ref{FIG-BASICA} - left).\footnote{A
typical example of an Ising spin glass is $\mathrm{Fe}_{0.5}
\mathrm{Mn}_{0.5} \mathrm{TiO}_3$.}

As we have said, the interaction between two given spins has a variable
sign. In some cases, neighboring spins may lower their energy if they are
parallel: their {\em coupling constant} $J_{ij}$, a number assigned to the
lattice link that joins spins $\sigma_i$ and $\sigma_j$, is then positive.
On the other hand, with roughly the same probability, it could happen that
the two neighboring spins prefer to anti-align (in this case,
$J_{ij}<0$).

A lattice link is {\em satisfied} if the two associated neighboring spins
are in the energetically favored configuration. In spin glasses, positive
and negative coupling constants occur with the same frequency, since the
spatial distribution of positive or negative $J_{ij}$ in the lattice links
is random; this causes {\em frustration}. Frustration means that it is
impossible to find an assignment for the spin values, $\sigma_i$, such
that all links are satisfied (the concept is sketched in
Fig.~\ref{FIG-BASICA} - right, and explained in the caption).  For any
closed lattice circuit such that the product of its links is negative, it
is impossible to find an assignment that satisfies every link. In spin
glasses, frustrated circuits usually arise with $50\%$ probability.

A longstanding model of this behavior is the Edwards-Anderson
spin glass~\cite{EDWARDS-ANDERSON}. The energy of a configuration (i.e., a 
particular assignment of all spin-values) is given by the Hamiltonian function
\begin{equation}
\label{eq:SG_ham}
H = - \sum_{\langle ij\rangle} \sigma_i J_{ij} \sigma_j\;,
\end{equation} where the angle brackets indicate that the  summation is
restricted to pairs of nearest neighbors in the lattice. The coupling constants
$J_{ij}$ are chosen randomly to be $\pm 1$ with $50\%$
probability\footnote{By saying $J=\pm1$ we are actually choosing our
energy units. For example, the critical temperature of $\mathrm{Fe}_{0.5}
\mathrm{Mn}_{0.5} \mathrm{TiO}_3$ is $T_c=20.7$ K, hence, taking into
account that in our units $T_c\simeq 1$, $J\sim \pm 3$ meV.}, and are kept
fixed. A given assignment of the $\{J_{ij}\}$  is called a {\em sample}.
Some of the physical properties (such as internal energy density, magnetic
susceptibility, etc.) do not depend on the particular choice for
$\{J_{ij}\}$ in the limit of large lattices (self-averaging property).
However, in the relatively small systems that one is able to simulate, it is
useful to average results over several samples.

It turns out that frustration makes it  hard to answer even the simplest
questions about the model.  For instance, finding the spin configuration
that minimizes the energy for a given set of $\{J_{ij}\}$ is an NP-hard
problem~\cite{barahona}.  In fact, our theoretical understanding of
spin-glass behavior is still largely restricted to the limit of high
spatial dimensions, where a rich picture emerges, with a wealth of
surprising connections to very different fields~\cite{MEPAVI}. In
particular, important clues on NP-completeness, including powerful
algorithms to solve hard K-clauses satisfiability problems have been
obtained from this analogy~\cite{ZECHINA}.

In three dimensions, we know experimentally~\cite{EXPERIMENTOTC} and from
simulations~\cite{SUE2000} that a spin-glass ordered phase is reached
below a critical temperature $T_\mathrm{c}$. In the cold phase
($T<T_\mathrm{c}$) spins {\em freeze} in some disordered pattern,
related to the configuration of minimal free energy.

For temperatures (not necessarily much) smaller than $T_\mathrm{c}$
spin dynamics becomes exceedingly slow.  In a typical experiment one
quickly cools a spin glass below $T_\mathrm{c}$, then waits to observe the
system evolution. As time goes on, the size of the domains where the spins
coherently order in the (unknown to us) spin-glass pattern, grows slowly. 
Domain growth is sluggish: even after eight hours of this process, for a
typical spin-glass material at a temperature $T=0.72 T_\mathrm{c}$, the
domain size is only around $40$ lattice spacings~\cite{SACLAY2004}.

The smallness of the spin-glass ordered domains precludes the experimental
study of equilibrium properties in spin glasses, as equilibration would
require a domain size of the order of $10^8$ lattice spacings. The good
news is that an opportunity window opens for numerical simulations.  In
fact, in order to understand experimental systems we only need to simulate
lattices significantly larger than the typical domain. This crucial
requirement has been met for the first time in the simulations discussed
herein.

\section{Monte Carlo simulations of spin glasses}\label{SG-MC-SECTION}

We evolve in Monte Carlo time the model of Eq. (\ref{eq:SG_ham}), defined on a
lattice of linear size $L$, by applying a Heat-Bath (HB)
algorithm~\cite{VICTORAMIT} that ensures that system configurations
$\mathcal{C}$ are sampled according to the Boltzmann probability distribution
\begin{equation} \label{eq:Boltz} P(\mathcal{C}) \propto
\exp\left({-\frac{H}{T}}\right) \ , \end{equation} describing the equilibrium
distribution of configurations of a system at constant temperature
$T=\beta^{-1}$. This is one well known Monte Carlo methods;
see e.g. \cite{Newman} for a review of other possible approaches; 

The local energy of a spin at site $k$  of a 3D lattice of linear size $L$
is \begin{equation} \label{eq:E_sigma} E(\sigma_k)=-\sigma_k
\sum_{m(k)}J_{km}\sigma_m\ \ , \end{equation} where the sum runs over the
six nearest neighbors, $m(k)$, of site $k$.

In the HB algorithm, one starts from the assumption that at any time any
spin has to be in thermal equilibrium with its surrounding environment,
meaning that the probability for a spin to take the value $+1$ or $-1$ is
determined only by its nearest neighbors. Following the Boltzmann
distribution, the probability for the spin to be $+1$ is
\begin{eqnarray}
P(\sigma_k=+1) & = &
\frac{e^{-E(\sigma_k=+1)/T}}{e^{-E(\sigma_k=+1)/T}+e^{-E(\sigma_k=-1)/T}}
=\frac{e^{\phi_k/T}}{e^{\phi_k/T}+e^{-\phi_k/T}} \ ,\label{eq:HB} \\ \label{eq:phi}
\phi_k & = & \sum_{m(k)} J_{km}\sigma_m \ , 
\label{eq:hbThings}
\end{eqnarray}

where $\phi_k$ is usually referred to as the \emph{local field} acting at site
$k$; note that$\phi_k$ takes only the $7$ even integer values in the range
$\left[-6,6\right]$.

The algorithm then reads as follows:
 \begin{enumerate}
 \item Pick one site
$k$ at random. 
 \item Compute the local field $\phi_k$ (Eq.~\ref{eq:phi}).
\item Assign to $\sigma_k$ the value $+1$ with probability
$P(\sigma_k=+1)$ as in Eq.~\ref{eq:HB}. This can be done by generating a
random number $r$, uniformly distributed in $[0,1]$, and setting $\sigma_k
= 1$ if $r < P(\sigma_k = 1)$, and   $\sigma_k = -1$ otherwise.
 \item Go
back to step 1.
\end{enumerate}
A full {Monte Carlo Step} (MCS) is usually defined as the iteration
of the above scheme for $L^3$ times.  By iterating many MCS, the
system evolves towards statistical equilibrium.

A further critically important tool for Monte Carlo simulations is Parallel
Tempering (PT)~\cite{PT1}. Dealing with spin glasses (but also proteins, neural
networks or several other complex systems) means handling rough free energy
landscapes and facing problems such as the stall of the system in a metastable
state. PT helps to overcome these problems  by simulating many  copies of the
system in parallel (hence the name) at different (inverse) temperatures
$\beta_i$ and allowing copies, whose $\beta$ (energy) difference is $\Delta
\beta$ ($\Delta E$), to exchange their temperatures with probability equal to
$\min\{1,\exp({\Delta\beta\Delta E}) \}$.  Typically one performs one PT step
every few  standard MCS.  During the PT dynamics, configurations wander from
the physically interesting low temperatures, where relaxation times can be long,
to higher temperatures, where equilibration is fast and barriers can be easily
traversed. Therefore, they explore the complex energy landscape more
efficiently, with correct statistical weights.  For an introduction to PT, see
for instance \cite{Kat-intro}.

We now sketch the steps needed to implement a Monte Carlo simulation on a
computer. First, note that  physical spin variables can be mapped to bits
by the following transformations
 \begin{align}
\sigma_k &\rightarrow S_k =  (1-\sigma_k)/2,\nonumber\\
J_{km} &\rightarrow \hat J_{km}  =  (1-J_{km})/2\
\label{eq:msc_map}
 \end{align}
We now define
\begin{equation}
F_k  =  \sum_{\langle km \rangle}\hat J_{km} \otimes S_m
\end{equation}
($\otimes$ represents the XOR operation). Each term in the sum for $\phi_k$
(Eq.  \ref{eq:hbThings}) is $\pm 1$ if the corresponding  $J_{km}$ and
$\sigma_m$ are equal or opposite, while in the sum for $F_k$  each $\hat J_{km}
\otimes S_m$ contributes $0$ if the two terms are equal or $1$ if they are
different; one then easily obtains (e.g., by explicit enumeration) a simple
relation between $\phi_k$ and $F_k$: $F_k = (6-\phi_k)/2$. In this way most
(but not all) steps in the algorithm involve logic (as opposed to arithmetic)
operations.

The following points are relevant:
 \begin{enumerate}
 \item High-quality
random numbers are necessary to avoid dangerous spurious spatial
correlations between lattice sites, as well as temporal correlations in
the sequence of generated spin configurations.
\item We already remarked that the local field
$\phi_k$ can take only the $7$ even integer values in the range
$\left[-6,6\right]$, so one computes probabilities $P(\sigma_k=+1)=f(\phi_k)$
once and stores them in a look-up table. 
\item The kernel
of the program is the computation of the local field $\phi_k$, involving
just a few arithmetic operations on discrete variable data. 
\item Under ergodicity and reversibility assumptions, the simulation
retains the desired properties even if the Monte Carlo steps are
implemented by visiting each lattice site exactly once, in any
deterministic order. Notice that the local field $\phi_k$ depends
only on the nearest neighbors of $k$.
 \item Several sets of couplings $\{J_{km}\}$ (i.e., \emph{different
samples}) need to be generated. An independent simulation has to be
performed for every sample, in order to generate properly averaged
results.
 \item One usually studies the properties of a spin-glass system by
comparing  the so-called \emph{overlaps} of two or more statistically
independent simulations of the
\emph{same} sample, starting from uncorrelated initial spin configurations
(the copies of a sample are usually referred to as \emph{replicas}).
\end{enumerate} 

The last three points above identify the parallelism available in the
computation; it can be trivially exposed for 5 and 6, while for 4 we need a
more accurate analysis. In fact, if we label all sites of the
lattice as \emph{black} or \emph{white} in a checkerboard scheme, all
black sites have their neighbors in the white site set, and \emph{vice
versa}: in principle, we can perform the steps of the algorithm on all
white or black sites in parallel. 

Our implementation of the HB algorithm on Janus leverages heavily on point 4
above, pushing to the extreme what is usually known as Synchronous Multi Spin
Coding (SMSC)~\cite{Newman}. SMSC was proposed to boost performance on
traditional CPUs; it maps spins of several lattice sites (no two of which are
nearest neighbors) onto corresponding bits of a large (e.g., 64 bit) data word
(this is referred to as multi-spin packing), so several steps of the algorithm
are performed in parallel as bit-wise logical operations. This technique speeds
up part of the algorithm, making random number generation (one random value is
needed for each spin update) the performance bottleneck. This bottleneck is
associated to a lack of logical resources and to memory bandwidth requirements
that cannot be satisfied even if most of the computational database resides on
cache. As shown later, on Janus we deploy several hundreds of update engines
(each handling one spin) and a matching number of random number generators,
ensuring that all needed memory accesses can be performed. 

We also mention that multi-spin packing may be arranged to exploit
``external'' parallelism, assigning spins of, for example, 64 or 128
\emph{independent} samples to each bit of the data word (dependent on word
size), and associating a data word to every lattice site.  This scheme is
usually known as \emph{asynchronous multi-spin coding} (AMSC). The
advantage of the AMSC is that the same random number can be used to update
all spins in a single data word (introducing a tolerable amount of
correlation). This approach has impressive benefits on traditional
architectures in terms of \emph{average update time per spin}; however it
does not shorten the \emph{wall-clock} time needed for a simulation, as it
does not improve (in fact, it worsens, with respect to SMSC) the time
needed to follow the history of one system. Also, it provides fewer and
fewer advantages as the system size increases, since single-system
simulation time increases, while the required number of samples 
decreases (due to the self-averaging property).

\section{Janus: the architecture}
\begin{figure} \begin{center}
\resizebox{0.38\columnwidth}{!}{\includegraphics{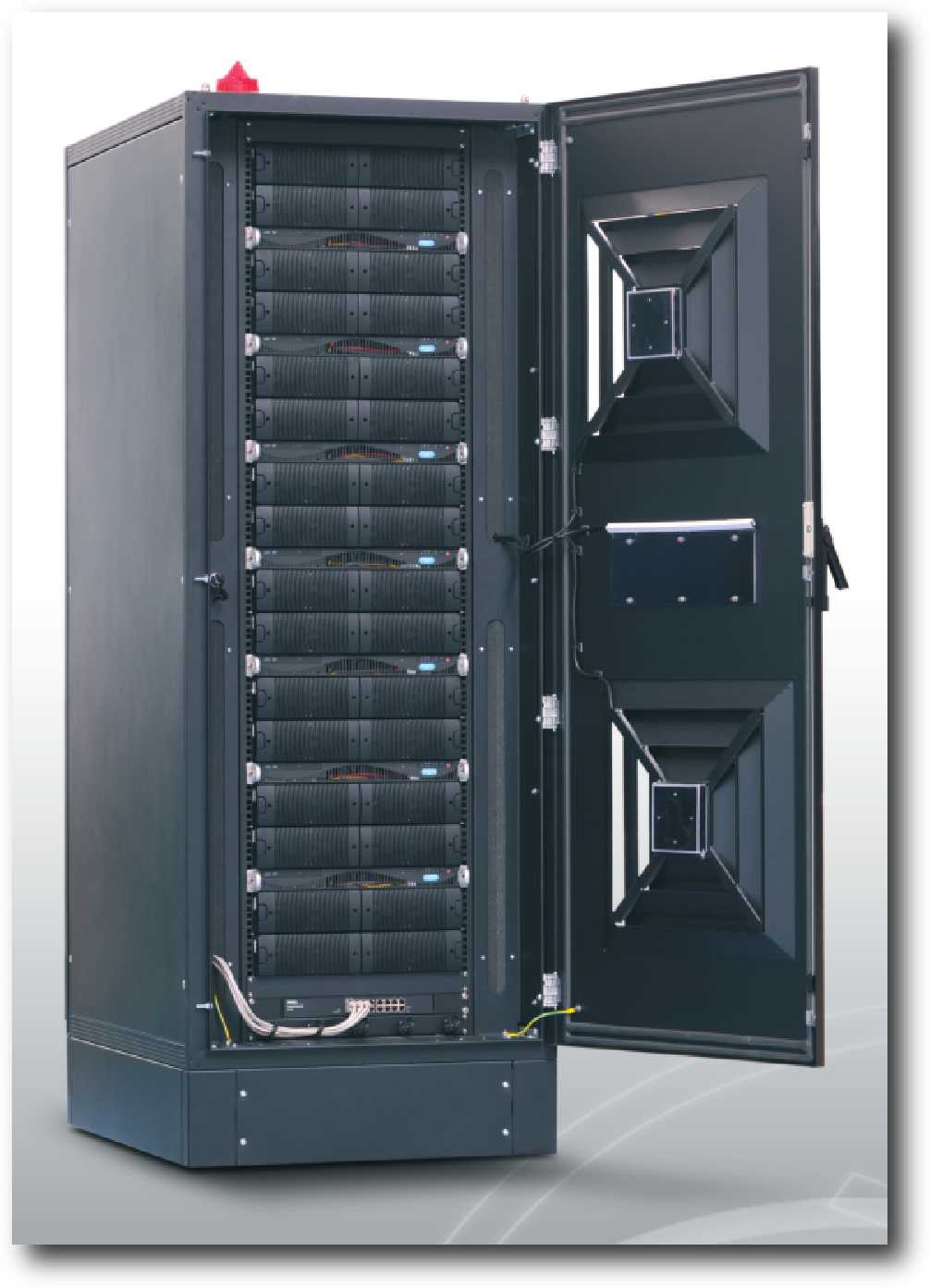} }
\resizebox{0.58\columnwidth}{!}{\includegraphics{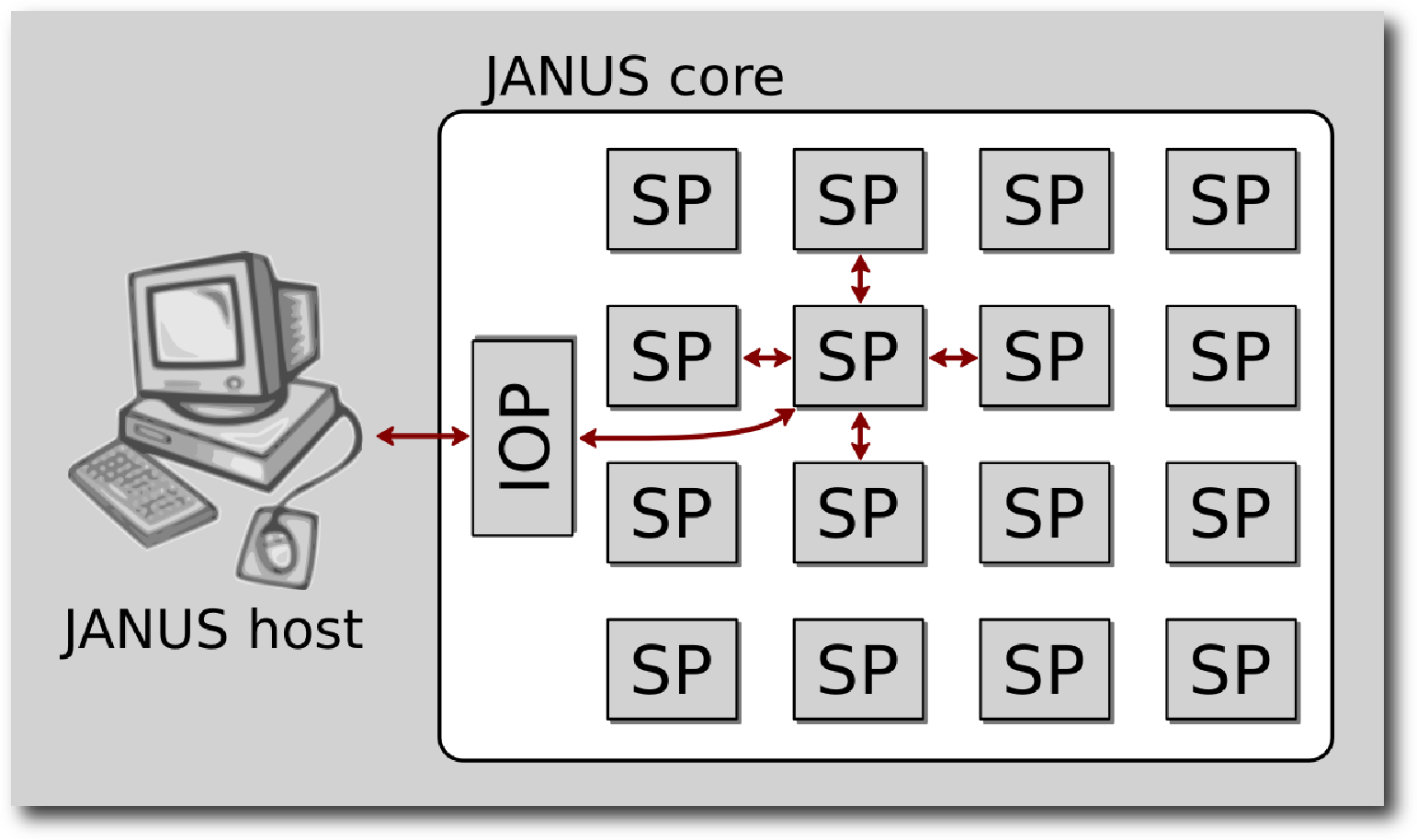} } \caption{{\bf
Left:} The Janus system used in this work: 16 Janus cores and 8
Janus hosts. {\bf Right:} Simple block diagram of a minimal Janus system,
with one Janus host and one Janus core.} \label{fig:architecture}
\end{center} \end{figure}

Janus is a heterogeneous massively parallel system built on a set of
FPGA-based reconfigurable computing cores, connected to a host system of
standard processors. Each Janus core contains  $16$ so-called simulation
processors (SP) and $1$ input/output processor (IOP). The SPs are
logically arranged as the nodes of a 2D mesh and have direct low-latency
high-bandwidth communication links with nearest neighbors. Each SP is also
directly connected with the IOP, while the latter communicates with an
element of the host system, as shown in the right side of
Fig.~\ref{fig:architecture}. SPs and IOPs are based on Xilinx
Virtex4-LX200 FPGAs. Our largest installation has 8 hosts (with a
4-TByte disk system) and 16 Janus cores (Fig.~\ref{fig:architecture},
left side).

The IOP handles the I/O interface, providing functionalities needed to
configure the SPs for any specific computational task, to move data sets
across the system and to control SP operation.  

The IOP processor exchanges data with the Janus host on two
gigabit Ethernet channels, running the standard raw-Ethernet communication
protocol. This is enough in our case, since most of the computation is
done within the SPs with little interaction with the host system.

The SP is the computational element of the Janus system. It contains
uncommitted logic that can be configured via the IOP as needed.  The SPs
of each Janus core can be  configured to run in parallel 16 independent
programs, or to run a single  program partitioned among the 16 FPGAs
exchanging data across the nearest-neighbor network.

Janus is programmed using two different codes: a standard C program which
includes a low-level communication library to exchange data to and from
the SPs, and a VHDL code defining the application running on the SPs. In
this way the FPGAs can be carefully configured and optimized for
application requirements. More details on the Janus architecture are given
in \cite{CPC,PARCO07,CISE06,CISE08}.

Tailoring our application to Janus has been a lengthy and complex
procedure, rewarded by huge performance gains. Obviously, a high-level
programming framework that would automatically split an application
between standard and reconfigurable processors and generate the corresponding
codes would be welcome.
Unfortunately the tools currently available do not deliver the needed
level of optimization and for Janus this work has to be done manually.

Our implementation of model and algorithm tries to exploit all internal
resources in the FPGA in a consistent way (see \cite{CPC} for a
detailed description). Our VHDL code is parametric in several key
variables, such as the lattice size and the number of parallel updates. In
the following description we consider, for definiteness, a lattice
of $80^3$ sites corresponding to the typical simulation described in the
introduction. 

LX200 FPGAs come with many small embedded RAM blocks, which may be
combined and stacked to naturally reproduce a 3D array of bits
representing the 3D spin lattice. In this example, we configure memory
blocks with  an 80-bit width to represent the linear size of the spin
lattice: 10 such memories with 10-bit addresses are enough to store our
whole lattice. Addressing all memories with the same given address $Z$
allows fetching or writing an $80\times 10$ portion of an entire $80^2$
lattice plane corresponding to spins with Cartesian coordinates ($0<x<79$,
$0<y<9$, $z$). The same portion on the next plane is addressed by $Z+80$,
referring to spins with coordinates ($0<x<79$, $0<y<9$, $z+1$), and so on.
Since we have to simulate two real replicas, we allocate one such
structure (the \emph{P} bank) for the \emph{black} spins of replica 1 and the \emph{white}
spins of replica 2.  We also allocate one similar structure (the \emph{Q}
bank) for white and
black spins of replicas 1 and 2 respectively. A similar storage strategy
applies to the read-only coupling bits. After initialization, the
machinery fetches all neighbor spins of an entire portion of plane of
spins and feeds them to the update logic. 

The update logic returns the processed (updated) portion of spins to be
uploaded to memory at the same given address.  Since we update $800$ spins
simultaneously, the update logic is made up by $800$ identical update
cells. Each cell must receive the 6 nearest-neighbors bits, the 6 coupling
bits and one 32-bit random number; it then computes the local field, which
is an address to a probability look-up table (LUT); 
look-up tables are small 32-bit wide memories instantiated as
\emph{distributed} RAM. There is one such table for each update cell.
The random number is then
compared to the probability value extracted from the LUT
and the updated spin is obtained.

Random number generators are implemented as 32-bit Parisi-Rapuano generators
\cite{PARISIRAPUANO}, requiring one sum and one bitwise XOR operation for each
number. The choice of a good enough random number generator is very critical
for large scale Monte Carlo simulation. We do not discuss this complex and
delicate point here, but rather refer the reader to an extended literature; for
the models in the Ising family (to which the Edwards-Anderson model belongs),
the Parisi-Rapuano generator has been analyzed from this point of view
in~\cite{victBall}.\footnote{We mention as well that, for $L\leq16$ lattices of
the Edwards-Anderson model, we performed very extensive comparisons with the
simulations of~\cite{CONTUCCI07} (performed on a PC, with a different random
number generator), finding an excellent statistical agreement. Furthermore, we
ran very long simulations for a single sample on a PC (using a combination of
64-bit Parisi-Rapuano and congruential PRNGs), which were then reproduced on
Janus. We obtained excellent agreement in the sample energy up to the PCs
precision of one part in $10^5$. This is a very strict test, because this
precision is considerably greater than that of any sample-averaged quantity
(the variation from sample to sample is much greater than the Monte Carlo error
within each sample).} To sustain the update rate, 10 generators, each made of
62 registers of 32 bits for the seeds, can produce 80 random numbers per clock
cycle each by combinatorial cascades.  The number of $800$ updates per clock
cycle is a good trade-off between the constraints of allowed RAM-block
configurations and available logic resources for the update machinery. 

Total resource occupation for the
design is $75\%$ of the RAM blocks (the total available RAM is $\simeq$
672 kB for each FPGA) and $85\%$ of the logic resources.  The system
runs at a conservative clock frequency of $62.5$ MHz. At this frequency,
the power consumption for each SP is $\simeq 35 W$. In the life time of the
project, we have further improved these codes; we are able
to update up to 1024 spins at each clock (this is
an upper limit as almost all logic resources available on the FPGA are used).

To perform the PT algorithm within a single FPGA node we can keep this
architecture almost unchanged.  The memory structure in this case
is just a generalization of the standard case. Since the values of
the couplings
do not change between the $N_T$ copies of the system at
different temperatures, the only additional data that we need to store
are the spin configurations of every copy that we
simulate. This is achieved by making the RAM blocks grow in depth and
using this extra space to store the $N_T$ spin configurations, as if we
were simulating a lattice of size $L^3\times N_{T}$ instead of $L^3$.

Information is stored in memories is a slightly different way w.r.t. the
previous case: remember that, when simulating only one temperature, we stored
the spin configurations of two replicas in memory banks P and Q, meshing their
black and white nodes; both replicas were simulated at the same temperature.  In
the PT case we want to simulate $N_{T}$ copies of the same system, each evolving
with a different $T$ value; in order to still use the configuration-meshing
trick seen before, we tangle together sites belonging to replicas at different
temperatures. Since we we need to simulate two (or more) replicas for each
temperature in order to obtain the required information about the overlap,  the
only solution in this case is to simulate exactly the same system and the same
set of temperatures independently.

For each temperature we need a different LUT for the Heat-Bath
probabilities, so we store on registers two different sets of values,
corresponding to the two temperatures being simulated in parallel. The update
cells are properly arranged so that each see either one or the other LUT. As
a consequence, the spins updated by each cell will be constantly working with
only one of the two available temperatures. This structure has no impact on
hardware complexity, since even in the non-PT case we replicate LUTs, so that
each LUT is read by only one update cell.

\begin{figure} \begin{center}
\resizebox{0.60\columnwidth}{!}{\includegraphics{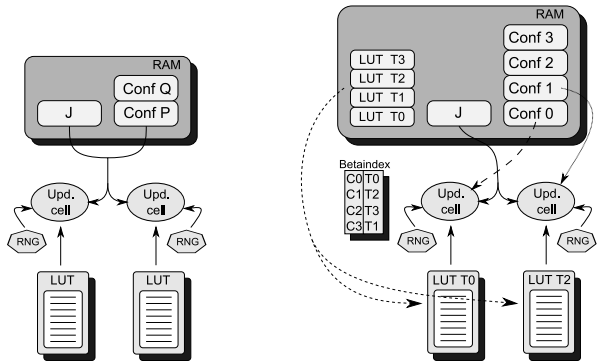} }
\caption{Firmware implementation with and without PT (simplified
representation). {\bf Left:} the standard algorithm with just one
temperature (i.e. a fixed LUT) and the relevant variables: couplings and
spin configurations. {\bf Right:} PT needs larger RAM space for all
LUTs and configurations, while the $J$ memory is left unchanged.
The BETAINDEX array controls the simulation temperature of each system.}
\label{fig:upd_nopt-pt} \end{center} \end{figure}

We store all temperature values and their corresponding pre-computed LUTs inside
the FPGA, in dedicated RAM blocks (see Fig. \ref{fig:upd_nopt-pt}).  The two LUT
sets needed by the actual simulation are copied into the LUT registers described
above.  An array, called BETAINDEX, keeps track of which system is being
simulated at which temperature. When two configurations have been completely
updated we move to another pair of systems, once again each characterized by a
(different) temperature value.  The BETAINDEX values of the configurations that
one is about to simulate point to the memory location where the corresponding
LUT values are stored. Once these have been loaded into the LUT register we are
ready to simulate the two new copies of the system.

The PT algorithm requires some new functions (and consequently new
hardware blocks) as well.  The first addition is related to the
computation (and storage) of the energy value of each configuration,
necessary to decide whether a temperature swap has to be done. This is
achieved by simply running the update algorithm over each replica, but
without actually updating the spin values in memories. A pipelined  binary
tree adder  sums up to 1024 local  energies in one cycle. Computing the
energy of a whole lattice takes exactly the same time as actually updating
the lattice. Once the energies of all configurations have been computed,
their values are used by the PT-engine.

The other important function must decide whether a temperature swap has to
be made. The PT algorithm works by comparing a random value with an
exponential, namely accepting the temperature swap when \begin{equation} r
\le e^{\Delta \beta \Delta E} \quad;\quad r \in [0,1) \ .  \end{equation}
In this case we cannot resort to the LUT trick used for the updates, since
it is impossible to pre-compute all values of $\Delta \beta \Delta E$.  On
the other hand, computing the exponential on the fly during the simulation
would dramatically slow down the performances of the machine.
We resort to the equivalent condition
\begin{equation} 
\ln r \le \Delta \beta \Delta E \quad;\quad r \in [0,1) \ ,
\end{equation}
exchanging the exponential function for the logarithm.  Computing the
logarithm is rather slow as well, but the argument of the logarithm does
not depend on the outcome of the MC step; we generate an appropriate
number of random values (using one 32-bit random generators) at the beginning
of the MC step and overlap the computation of the logarithms with the
actual simulation.  Once the MC steps at all temperatures are completed,
we just have to evaluate the products $\Delta E ~\Delta \beta$ and compare
them with the $\ln r$ to decide whether to accept
or reject the PT switch.  In the former case we exchange the index values
within the BETAINDEX array and move to the next pair of neighboring
temperatures.  Once all  temperature pairs have been evaluated we are
ready to start again with the simulations, with a reshuffled temperature
order.

The PT implementation described in this section is our best choice, for
small lattice sizes, because it is self-contained (within a single SP) and
does not need too much time-wasting communications.  On the other hand,
due to limited FPGA memory resources, this implementation works only with
smaller lattice sizes ($L \le 32$) and a limited number of temperatures
($N_{T} \le 128$). 

\section{Simulation details and physics results} \label{sect:results-Phys}
In almost 4 years of continuous operation, Janus has been used to study in
detail several important aspects of the physics of spin glasses. We have
studied the Ising spin glass introduced earlier in the paper as well as
the Potts glass model, in which the spin variables are still discrete, but
no longer binary. We present a summary of our results in the following
sub-sections; full details are available in the corresponding original
papers.

\subsection{Ising Spin Glasses}\label{sect:results-IEA}
Janus started production in spring 2008 with a study of the off-equilibrium
dynamics in the three-dimensional Edwards-Anderson (EA) model with binary
couplings (Eq.~\ref{eq:SG_ham}). We simulated a large lattice ($L=80$)
using HB dynamics for more than $10^{11}$ Monte Carlo steps
(MCS), both at the critical point and at some lower temperatures. This
roughly corresponds to 0.1 seconds on a real spin glass, orders of
magnitude longer than any previous simulation and already long enough to
understand what happens at the experimentally relevant scale.
Our results are described in detail in \cite{Janus3DEA_PRL0,Janus3DEA_JSP}.

Subsequently, we ran a very long simulation campaign for the study of
low-temperature equilibrium properties. We obtained several equilibrium
configurations for 3D systems of size up to $L=32$ for many samples
($4000$ disorder realizations for $L=16,24$ and $1000$ for $L=32$) at
temperatures as low as $\approx 64\%$ of the critical temperature, even
for the largest system size.  In order to increase the statistics of
overlap-dependent observables and to avoid bias problems in four-point
correlation functions, we simulated four independent replicas per sample.
We used the PT update scheme, interleaved with HB dynamics.

One important technical innovation of these simulations (expanding on a
method first introduced in~\cite{HEISENBERG}) was the use of a strict
thermalization criterion based on the temperature random walk during the
PT. The \emph{synchronous} structure of the update algorithm available on
Janus made it possible to tailor the simulation schedule of every single
sample, ensuring that longer run-time was available to those disorder
realizations for which it was harder to meet our thermalization criteria.
In fact, a handful of samples of size $L=24$ and $L=32$ needed up to six
months wall-clock time of simulation.   Our first results at equilibrium
were published in~\cite{Janus3DEA_JSM} (see also this reference for
details on our simulation parameters and thermalization criteria).  Since
then, we have performed more specialized analyses to expand our
understanding of the equilibrium spin-glass phase~\cite{Janus3DEA_PRB,TCJ}
as well as to study its relationship to the (experimentally relevant)
non-equilibrium evolution~\cite{Janus3DEA_PRL}.

In general terms, our main interest was to understand the low-temperature
properties of the spin glass in order to check predictions by widespread
theories, such as Replica Symmetry Breaking~\cite{SomeRSBpaper} (RSB), the
Droplet Model~\cite{SomeDMpaper} (DM) and the
Trivial-Not-Trivial~\cite{SomeTNTpaper} picture (TNT).  We addressed, at
equilibrium and out-of-equilibrium, the following problems: aging and full
aging in $C(t,t_\text{w})$; spatial correlations, coherence length, and
dynamical critical exponent $z(T)$; the statics-dynamics equivalence and
finite-time scaling; dynamical heterogeneities and the phase transition at
$q=q_\mathrm{EA}$; equivalence among the different definitions of the
overlap; the sample-to-sample fluctuations of the probability distribution
of the overlap and the finite-size scaling of pseudocritical temperatures.
In the following we give an outline of some of the main lines of
this research.

We begin with the non-equilibrium study. We consider a system starting
with a random configuration (equivalent to infinite temperature) that is
instantly cooled to a working temperature $T<T_\text{c}$, left to
evolve for a waiting time $t_\text{w}$ and finally investigated at a time
$t+t_\text{w}$. We consider the following correlation functions:
\begin{align} 
C(t,t_\text{w})&=\overline{\frac{1}{N} \sum_x \sigma_x(t_\text{w})
\sigma_x(t+t_\text{w})}\,,\\
C_4(r,t_\text{w})&=\overline{\frac{1}{N} \sum_x \sigma_x(t_\text{w}) \tau_x(t_\text{w})
\sigma_{x+r}(t_\text{w}) \tau_{x+r}(t_\text{w})}\,,\\
C_{2+2}(r,t,t_\text{w})&=\overline{\frac{1}{N} \sum_x \sigma_x(t_\text{w})
\sigma_x(t+t_\text{w}) \sigma_{x+r}(t_\text{w}) \sigma_{x+r}(t+t_\text{w})- C^2(t,t_\text{w})}\,,
\end{align}
where $\sigma$ and $\tau$ denote spins in independent
replicas of the system.

$C(t,t_\text{w})$ is the most basic indicator of the correlation in the
system, it simply gives the ``memory'' at $t+t_\text{w}$ of the
configuration at $t_\text{w}$. In particular,  notice that
\begin{align}
\lim_{t_\text{w}\to\infty} \lim_{t\to\infty} C(t,t_\text{w}) &= 0,&
\lim_{t\to\infty} \lim_{t_\text{w}\to\infty} C(t,t_\text{w}) &=
q_\text{EA},
\end{align}
where $q_\text{EA}$ is the spin-glass order parameter.  

The second function, $C_4(r,t_\text{w})$, gives the spatial correlation in the
system. We have used it to study the size of the growing coherent domains,
through the computation of a coherence  length $\xi(T,t_\text{w})$.  We computed
the dynamical critical exponent $z(T)$ using $\xi(T,t_\text{w}) \propto
t_\text{w}^{1/z(T)}$, checking that $z(T)$ follows roughly the law $z(T)=z(T_c)
T_c/T$. Extrapolating $\xi(t_\text{w})$ to the experimental times we found a good
agreement with experiments. In addition, we characterized with high
precision the behavior of $C_4(r,t_\text{w})$, allowing us to recover the
anomalous dimension of the overlap and the replicon exponent.

We also found that the dynamics is heterogeneous, with spatial regions
behaving in different ways. This phenomenon cannot be studied using just
$C_4(r,t)$ or $C(t,t_\text{w})$, so we resorted to $C_{2+2}(r,t_\text{w})$
and computed its associated correlation length, $\zeta(t,t_\text{w})$.
Taking advantage of the fact that $C(t,t_\text{w})$ is monotonic for fixed
$t_\text{w}$, we studied this observable as a function of $t_\text{w}$ and
$C(t,t_\text{w})$ and found that it experienced a crossover.  For small
$C$, $\zeta(t,t_\text{w})$ diverges at large $t_\text{w}$ just like the
coherence length $\xi(t_\text{w})$ does; while it tends to a
$t_\text{w}$-independent value at large correlations. 
In order to take this study to a completely quantitative level, we found
that non-equilibrium data was not enough. 

Fortunately, we were able to draw on the deep connection between the
(experimentally unreachable) equilibrium phase and the non-equilibrium
dynamics. Indeed, we found that a quantitative time-length
dictionary could be established, relating data in the thermodynamic limit
for finite time $t_\text{w}$ with finite lattices at equilibrium for some
size $L(t\text{w})\propto \xi(t_\text{w})$ (the proportionality factor
depends on the precise definition of $\xi$). If this time-length dictionary
is followed, then one can define equilibrium analogues for each of the
off-equilibrium correlation functions seen above and obtain a quantitative
matching between both sets of data.\footnote{We can motivate this
equivalence naively by considering  the
infinite system at $t_\text{w}$ as composed of many equilibrium
subsystems of size $\xi(t_\text{w})$.}

Using this equivalence, we found that the crossover in the dynamical
heterogeneities is caused by the onset of an actual phase
transition at $q=q_\text{EA}$. We were then able to explain the
non-equilibrium evolution with a finite-time scaling ansatz, using
parameters computed in equilibrium. In fact, our study of the equilibrium
connected correlation functions allowed us to obtain the first reliable
estimate of $q_\text{EA}$ itself. 

The fact that finite-size scaling at equilibrium and finite-time scaling
in the dynamics are strictly related through this time-length dictionary
has strong experimental implications: the time scale of experiments on
real samples are roughly $1$ hour, corresponding to about $4\times
10^{15}$ MC steps. This translates to a equivalent equilibrium system size
of $L\simeq 110$.

We found that in this large (but finite) limit overlap
equivalence, i.e., the completeness of the description of the
low-temperature phase by a single overlap definition, holds for both our
equilibrium and out-of-equilibrium data. It is worth noting that overlap
equivalence is required by RSB, while the TNT scenario predicts
that, for instance, the link overlap $Q_\mathrm{link}$ would be
uncorrelated to the spin overlap $q$.

This discussion notwithstanding, we also considered extrapolations to
the thermodynamical limit. In this infinite-size limit we found that the
finite-size corrections needed to fulfill some DM predictions would imply
a non-physical extrapolation of the Edwards-Anderson order parameter
$q_\mathrm{EA}$ and that the overlap probability density functions $P(q)$
for our Monte Carlo data have a non-vanishing tail down to $q=0$.

An analysis of the sample-to-sample fluctuations of the $P(q)$ shows that our
data are compatible with replica equivalence (one-replica observables are
replica-symmetric also in the RSB solution) ---or equivalently stochastic
stability (system stability upon small long-range perturbations in the
Hamiltonian)--- and that triangular relations predicted by RSB are satisfied
with increasing accuracy as the system size increases.  Finite-size overlap
distributions are in agreement with the ones obtained by introducing controlled
finite-size effects to the mean-field predictions.

Finally, we have recently completed a large-scale simulation campaign of the
$D=4$ Edwards-Anderson Ising spin glass with a non-vanishing applied magnetic
field~$h$ in order to study the stability of the spin-glass phase. According to
the RSB picture (which we know is valid above the upper critical dimension
$D=6$) the spin-glass phase should survive the application of a magnetic field,
leading to the de Almeida-Thouless line of phase transitions~\cite{DAT}. The
Droplet Model, on the other hand, expects that even an infinitesimal  magnetic
field will destroy the spin-glass phase. In our study~\cite{DAT-PNAS}, the large
systems simulated and the use of novel finite-size scaling techniques have
allowed us to find clear signs of a second-order phase transition and to
characterize its critical exponents.

\subsection{Potts glass}\label{sect:results-PG}

We have also used Janus to simulate the Potts glass model, in which each spin
$\sigma_i$ can take $p$ values. The Hamiltonian of the Potts glass
model reads \begin{equation} \mathcal{H}=-\sum_{\langle i,j \rangle}
J_{ij} \delta(\sigma_i,\sigma_j)\,, \end{equation}  where $\delta$ is the
Kronecker delta function and $J_{ij}$ are the bimodal couplings between the two
spins ($J_{ij}=\pm 1$ with equal probability). For a mean-field (MF) analysis of the
model, see for example Ref. \cite{GROSSKANTERSOMPOLINSKI}. 

Our main aim was to characterize how the critical behavior depends on the
value of $p$.  We studied the Potts glass with $p=4,5$ and $6$, and
various linear sizes up to $L=16$ (see
Refs.~\cite{JANUSPOTTS_4,JANUSPOTTS_56}). $L < 8$ lattices were simulated
on conventional computers while the larger ones ($L=8,12,16$) were
simulated on Janus. Even using Janus we were unable to thermalize a
significant set of samples (for $p\geq5,\;L=16$) due to the very long time
required. In particular it has become clear that simulations severely slow
down with increasing $p$. We simulated several thousand samples for each
($p,L$) pair to analyze the critical behavior, evaluate the critical
exponents and measure the critical temperature. It was also absolutely
mandatory to study the possibility of the onset of ferromagnetic behavior.
Systems were simulated up to $10^{10}$ MCS, using PT and the HB algorithm
for local updates, using the same strict thermalization criteria as in the
Edwards-Anderson model (based on the study of the temperature random walk).

Our results, based on the scaling analysis of the second-moment
correlation length, indicate that, although MF predicts a change from a
second-order to a first-order transition for $p>4$, in the
three-dimensional model this change does not happen. The thermal critical
exponent $\nu$ drifts toward the lower bound admissible for a second
order phase transition ($2/3$ for a 3D system, see e.g.
\cite{ESCALA-DES}), but does not cross it. We also obtained a
simple relationship between the critical inverse temperature and the
number of states, $\beta_c \approx p$. Also, we did not find, for all
simulated temperatures, any ferromagnetic effects, which are allowed in the
model and predicted to be relevant at low-enough temperatures in MF
theory. This has been analyzed both by determining the ferromagnetic
anomalous dimension ($\eta_m$) and by studying $\chi_m$ and $\langle
\overline{|m|} \rangle$.

\section{Janus performance} \label{sec:perfs}

In this section we analyze the performance of the Janus computer for some
of the applications described in Sect.~\ref{sect:results-Phys} and
compare with that of commodity systems based on more traditional computer
systems, available both when Janus was developed as well as today.

At the time the Janus project started, early 2006, state-of-the-art
commodity systems where based on dual-core CPUs. Prior to actually
building the system we made an extensive analysis of the performance gain
that we could expect from the new machine (see for instance \cite{CPC}).
Table ~\ref{tab:tableperformance} contains a short summary of that
analysis, listing the relative speed-up for the Ising and the Potts
models of one Janus SP with respect to standard processors available in
2006-2007 .

\begin{table} \begin{center} 
\begin{tabular}{|c|c|c|c|} \hline Model & Algorithm &  Intel Core 2 Duo &
Intel i7 \\ \hline 3D Ising EA & Metropolis & $45\times$ & $10\times$ \\ \hline 3D Ising
EA & Heat Bath  & $60\times$ & - \\ \hline p=4 3D glassy Potts & Metropolis  &
$1250\times$ & - \\ \hline \end{tabular}
 \caption{Speed-up factors of one Janus
SP with respect to state-of-the-art CPUs available at the time the project
was started. \label{tab:tableperformance}}  \end{center}
\end{table} 

Since the deployment of Janus, in  Spring 2008, significant improvements have
been made in the architecture and performance of commodity architectures,
and in spite of that, Janus is still a very performing machine.    

We have extensively compared ~\cite{PPAM09,PARA10} Janus with several
multi-core systems based on the IBM Cell Broadband Engine, the multi-core
Nehalem Intel CPU, and the NVIDIA Tesla C1060 GP-GPU. We have made this
exercise for the Ising model (as opposed to the Potts model) as in the
former case the relative speed-up is much smaller, so we may expect
traditional processors to catch up earlier.
We consider these results as state-of-the-art comparisons, assuming that
within a factor 2 they are still valid for even more recent multi-core
architectures, like the Fermi GPUs. This assumption is indeed verified by
an explicit test made on the very recent 8-core Intel Sandy Bridge
processor.
  
As discussed in previous sections, for traditional processor architectures
we analyzed both SMSC and AMSC strategies and also considered mixes of the
two tricks (e.g., simulating at the same time $k$ spins belonging to $k'$
independent samples), trying to find the best  option from the point of
view of performance.

A key advantage of Janus is indeed that there is no need to look for these
compromises: an SP on Janus is simply an extreme case of SMSC
parallelization: if many samples are needed on physics ground, more SPs
are used. Equally important, if different samples need different numbers
of Monte Carlo sweeps (e.g., to reach thermalization),  the length of each
simulation can be individually tailored without wasting computing
resources on other samples.

Performance results for Janus are simply stated: one SP updates $\approx
1000$ spins at each clock cycle (of period $16$ ns), so the spin update
time is $16$ ps/spin for any lattice size that fits available memory.
For standard processors, we collect our main results for the 3D Ising
spin glass in tables \ref{SUTEA} and \ref{GUTEA} for SMSC and AMSC
respectively.

\begin{table}[t] \centering \scalebox{0.85}{
\begin{tabular}{||c|c|c|c|c|c|c||} \hline \multicolumn{7}{|c|}{3D Ising
spin-glass model, SMSC (ns/spin)} \\ \hline \hline L   & Janus SP & I-NH
(8-Cores) & CBE (8-SPE) & CBE (16-SPE) & Tesla C1060 & I-SB (16 cores) \\
\hline 16  & 0.016  & 0.98           &  0.83       &  1.17        & --
& --              \\ 32  & 0.016  & 0.26           &  0.40       &  0.26
& 1.24        & 0.37            \\ 48  & 0.016  & 0.34           &  0.48
&  0.25        & 1.10        & 0.23            \\ 64  & 0.016  & 0.20
&  0.29       &  0.15        & 0.72        & 0.12            \\ 80  &
0.016  & 0.34           &  0.82       &  1.03        & 0.88        & 0.17
\\ 96  & --     & 0.20           &  0.42       &  0.41        & 0.86
& 0.09            \\ 128 & --     & 0.20           &  0.24       &  0.12
& 0.64        & 0.09            \\ \hline \end{tabular} } \caption{SMSC
update time for the 3D Ising spin-glass (binary) model, for Janus and for
several state-of-the-art processor architectures.  I-NH (8-Cores) a
dual-socket quad-core Intel Nehalem board, CBE (16-SPE) is a dual-socket
IBM Cell board, and I-SB a dual-socket eight-core Intel Sandy Bridge
board. \label{SUTEA}} 
\scalebox{0.85}{
\begin{tabular}{||c|c|c|c|c|c|c||} \hline \multicolumn{7}{|c|}{3D Ising
spin-glass model, AMSC (ns/spin)} \\ \hline \hline L   & Janus      & I-NH
(8-Cores) & CBE (8-SPE)  & CBE (16-SPE) & Tesla C1060 & I-SB (16 cores) \\
\hline 16  & 0.001 (16) & 0.031 (32)     &  0.052 (16)  &  0.073 (16)  &
--          & --		\\ 32  & 0.001 (16) & 0.032 ( 8)     &
0.050 ( 8)  &  0.032 ( 8)  & 0.31 (4)    & 0.048 ( 8)	\\ 48  & 0.001
(16) & 0.021 (16)     &  0.030 ( 8)  &  0.016 (16)  & 0.27 (4)    & 0.015
(16)	\\ 64  & 0.001 (16) & 0.025 ( 8)     &  0.072 ( 4)  &  0.037 ( 4)
& 0.18 (4)    & 0.015 ( 8)	\\ 80  & 0.001 (16) & 0.021 (16)     &
0.051 (16)  &  0.064 (16)  & 0.22 (4)    & 0.011 (16)      \\ 96  & --
& 0.025 ( 8)     &  0.052 ( 8)  &  0.051 ( 8)  & 0.21 (4)    & 0.012 ( 8)
\\ 128 & --         & 0.025 ( 8)     &  0.120 ( 2)  &  0.060 ( 2)  & 0.16
(4)    & 0.011 ( 8)	\\ \hline \end{tabular} } \caption{AMSC update time for
the 3D Ising spin-glass (binary) model, for the same systems as in the
previous table. For Janus, we consider one core with 16 SPs. The
number of systems simulated in  parallel in the multi-spin approach is
shown in parentheses} \label{GUTEA} \end{table}

We see that performance (weakly) depends also on the size of the simulated
lattice: this is an effect of memory allocation issues and of cache
performance.  All in all, recent many-core processors perform today much
better that 5 years ago: the performance advantage of Janus has declined
by a factor of approximately $10$ for SMSC: today one Janus SP outperforms
very latest generation processors by just a factor $5\times \cdots
10\times$. It is interesting to remark that GP-GPUs are not the most
efficient engine for the Monte Carlo simulation of the Ising model: this
is so, because GP-GPU strongly focus on floating-point performance which
is not at all relevant for this specific problem.  There is one point
where Janus starts to show performance limits, it is associated to the
largest system size that the machine is able to simulate: no significant
limit applies here for traditional processor.

All in all, for the specific applications we have presented in this paper,
Janus ---after 4 years of operation--- still has an edge of approximately
one order of magnitude, which directly translates on the wall-clock time of
a given simulation campaign.

One may try an approximate assessment of the potential performance of a
system based on the Janus architecture using state-of-the-art technology
available in mid 2012. The breakup of the individual factors is shown in
Table \ref{Janus2}, based on some exploratory tests made with Xilinx
Virtex-7 FPGAs. We see that one FPGA-based SP increases performance by a
factor of order $10\times$. For a new generation machine it would be
important to split the simulation of one sample over all SPs available on
one Janus core, enabling a further $16\times$ SMSC performance gain. This
implies that a fast enough interconnection structure has to be developed.

\section{Conclusions}
This paper has described in detail the Janus computer architecture and how
we have configured the FPGA hardware to simulate spin-glass models on this
architecture. We have also briefly reviewed the main physics results that
we have obtained in approximately 4 years operating with this machine.

From the point of view of performance, Janus  still has an edge on
computing systems based on state-of-the-art processors, in spite of the
huge architectural developments since the project was started.  It is
certainly possible to reach very high performances in terms of spin flips
per second using multi-spin coding on CPUs or GP-GPUs (or simply by
spending money on more computers), thus concurrently updating many samples
and achieving very large statistics. In the Janus collaboration, we have
instead concentrated on a different performance goal: minimizing the
wall-clock for a very long simulation, by concentrating the updating power
in a single sample. This has allowed us to bridge the gap between
simulations and experiments for the non-equilibrium spin-glass dynamics or
to thermalize large systems at low temperatures, thus gaining access to
new physics. In particular, one single SP of Janus is able to simulate
(two replicas of) an $L=80$ three-dimensional lattice for $10^{11}$ MCS in
about 25 days.

Janus is one of the few examples of the development of a successful
computing environment fully based on reconfigurable computing elements.
This success comes at the price of a large investment in mapping and
optimizing the application programs onto the reconfigurable hardware. This
has been possible in this case as the Janus group has a full
understanding of all facets of the algorithms and every performance gain
immediately brings very large dividends in term of a broader physics
program.

Most potential FPGA-based applications do not have equally favorable
boundary conditions, so automatic mapping tools would be most welcome;
however their performance is still not enough for a widespread use of
FPGA-based computing. Nevertheless, there is still room for substantial
progress, even in the field of spin glasses. Nowadays, the theoretical
analysis of temperature-cycling experiments is still on its infancy. Janus
has made possible an in-depth investigation of isothermal aging (i.e.,
experiments where the working temperature is kept constant). However,
isothermal aging reflects only a minor part of the experimental work,
where different temperature variation protocols are used as a rich probe
of the spin-glass phase. Memory limitations have proven to be a major
problem for Janus, because the coherence length grows very fast close to
the critical temperature (the simulated system must be, as a rule of
thumb, at least eight times larger than the largest coherence length
attained during the simulation). A new generation machine, Janus II,
expected to enter full operation by the end of 2012, should provide a
major asset to overcome this limitation. According to the estimates at the
end of Sect. 6, we should be able to reach the same time scales of
$10^{11}$ lattice sweeps ---which is roughly equivalent to a tenth of a
second--- on systems containing some $5\times10^7$ spins. In other words, we
should be able to simulate systems with lattice sizes up to $L=400$, large
enough to accommodate a coherence length of up to 50 lattice spacings.
After 40 years of investigations, a direct comparison between experiments
and the Edwards-Anderson model will be finally possible.

\begin{table}[t] 
\centering \scalebox{0.85}
{ \begin{tabular}{||l | r||} \hline 
Feature   & Factor \\ \hline
FPGA size       & $2.5 \cdots 4$x \\
Clock frequency & $4.0$x \\ 
Core-parallel   & $16$x\\
Grand Total     & $160 \cdots 200$x\\  \hline 
\end{tabular} }
\caption{Performance factors for a tentative re-engineering of Janus using
state-of-the-art technology
available in mid 2012.}
\label{Janus2}
\end{table}

\section*{Acknowledgments}
We wish to thank several past members of the Janus Collaboration, F.  Belletti,
M. Cotallo, D. Sciretti and J.L. Velasco, for their important contributions to
the project. Over the years, the Janus project has been supported by the EU
(FEDER funds, No.  UNZA05-33-003, MEC-DGA, Spain), by the MICINN (Spain)
(contracts FIS2006-08533, FIS2009-12648, FIS2007-60977, FIS2010-16587,
FPA2004-02602, TEC2010-19207), by CAM(Spain), by the Junta de Extremadura
(GR10158), by UCM-Banco Santander (GR32/10-A/910383), by the Universidad de
Extremadura (ACCVII-08) and by the Microsoft Prize 2007. We thank ETHlab for
their technical help.  E.M. was supported by the DREAM SEED project and by the
Computational Platform of IIT (Italy). M.B.-J. and B.S.  were supported by the
FPU program (Ministerio de Educaci\'on, Spain); R.A.B.  and J.M.-G. were
supported by the FPI program (Diputaci\'on de Arag\'on, Spain); finally
J.M.G.-N. was supported by the FPI program (Ministerio de Ciencia e
Innovaci\'on, Spain).

\end{document}